\documentstyle[aps,preprint,multicol,epsf]{revtex}
\newcommand{\gl}[1]{(\ref{#1})}

\begin{document}

\title{Structure and structure relaxation}

\author{T.~Franosch, W.~G{\"o}tze, M.~R.~Mayr, and A.P.~Singh}

\address{Physik--Department, Technische Universit{\"a}t
M{\"u}nchen, 85747
Garching, Germany}

\smallskip

\date{J. Non-Cryst. Solids, in print}

\maketitle

\begin{abstract}
\noindent
A discrete--dynamics model, which is specified solely in terms of the
system's equilibrium structure,  is defined for the density correlators
of a simple fluid. 
This model yields results for the
evolution of glassy dynamics which are identical with the ones
obtained from the mode--coupling theory for ideal liquid--glass
transitions. The decay of density fluctuations outside the
transient regime is shown to be given by a superposition of Debye
processes. The concept of structural relaxation is given a precise meaning.
It is proven that the long--time part of the
mode--coupling--theory solutions is structural relaxation, while the transient
motion merely determines an overall time scale for the glassy dynamics.
\bigskip


\medskip

\end{abstract}

\vspace{2cm}

\newpage

%\begin{multicols}{2}
\section{Introduction}

Glass--forming liquids exhibit a dynamics which appears anomalous
in comparison to the one of conventional condensed matter. The
characteristic time scale $\tau$ for the motion can be several
orders of magnitude larger than the natural time scale  of, say, normal liquid dynamics. The
scale $\tau$ is extremely sensitive to changes of control
parameters such as temperature $T$ or density $n$; for example, a change of $T$
by 10 degrees may imply a change of $\tau$ by a factor 100.
Furthermore, decay of correlations or spectra are stretched over
dynamical windows of time $t$ or frequency $\omega$, respectively, 
of many decades in size \cite{Wong76}. The evolution of the
anomalous dynamics upon cooling or compressing a liquid has first been
studied comprehensively by Li {\it et al.} \cite{Li92,Cummins93} for the mixed salt
${\rm Ca}_{0.4} {\rm K}_{0.6} ({\rm NO}_3)_{1.4}$ (CKN) and by van Megen and
Underwood for a
hard--sphere colloid \cite{Megen94b}. The former determined depolarized--light--scattering 
spectra for a four--decade frequency window  and the
latter measured photon--correlation curves for an eight--decade
time window. In recent years the mode--coupling theory (MCT) for
the glass transition has been developed \cite{Goetze92}. It deals with a
mathematically well--defined model for anomalous dynamics, which
results from a bifurcation point; its novel features are due to
the interplay of non--linearities with divergent retardation
times. Results of this theory have been tested against
experiments, for example in Refs. \cite{Li92,Cummins93,Megen94b}. The outcome of these
tests qualifies MCT as a candidate for a theory of the anomalous  dynamics in
glass--forming liquids.

The equilibrium structure of a classical system is determined by
the ratio of interaction potentials and thermal energy via the
Boltzmann factors. It is independent of the particle masses or
other inertia parameters. It is the same for a conventional
system, ruled by Newton's equations of motion, and for a colloid,
whose time evolution is controlled by a Brownian dynamics. The
simplest information on structure is provided by the structure
factor $S_q \propto \langle | \varrho_{\vec q}|^2 \rangle$,
where $\varrho_{\vec q}$ are density fluctuations of wave vector
${\vec q}$; here and in the following $ q = | {\vec q}|$ denotes the vector modulus,
and $\langle \, \rangle$ abbreviates canonical averaging.
The indicated anomalous dynamics in glassy
systems 
 is not related to anomalies of the equilibrium
structure; $S_q$ is a smooth function of $n, T$ and $q$
throughout the
whole liquid regime. The simplest information on structure
dynamics is provided by the density correlators $\Phi_q(t) =
\langle \varrho_{\vec q} (t)^* \varrho_{\vec q} \rangle / \langle |
\varrho_{\vec q}|^2 \rangle$. The evaluation of these functions
is the main theme of MCT. In this paper we will restrict
ourselves to  simple liquids and to the idealized version of
the MCT \cite{Goetze92}.

Two propositions shall be considered. First, the anomalous dynamics
is solely determined by the equilibrium structure, i.e., by the
potential landscape in the  configuration space.
Second, the anomalous dynamics can be described by a
superposition of Debye--relaxation processes:
\begin{equation}
\label{A1}
\Phi_q (t) = \sum_j \rho_{q,j} \exp (- \gamma_{q,j}t) \,\, ,
\quad \rho_{q,j} > 0 \,\, , 
\quad \gamma_{q,j} \geq 0 \,\, .
\end{equation}
\noindent These propositions provide a precise meaning  to the statement that 
the
anomalous dynamics is  structure relaxation. 

Density correlators in
conventional liquids exhibit the short--time expansion $\Phi_q (t)
= 1- \frac{1}{2} (\Omega_q t)^2 + O (t^3)$, where $\Omega_q^2 =
v^2 q^2/S_q$, 
with $v^2 = (k_B T/m)$ denotes the bare phonon frequency \cite{Hansen86}.
This expansion contradicts Eq. \gl{A1}. Furthermore, the thermal velocity $v$
depends on the particle mass $m$ and is therefore not an
equilibrium quantity. It is necessary to formulate the
propositions more precisely to avoid a conflict with the cited
short--time behavior. Thereby some insight in the physics of
glassy dynamics shall be provided.

\section {Basic equations}\label{sectionzwei}

Within the Zwanzig--Mori formalism one can derive the equation
\begin{equation}
\label{A2}
\partial_t^2 \Phi_q (t) + \Omega_q^2 \Phi_q (t) +
\int_0^t M_q (t-t') \partial_{t'} \Phi_q (t') {\rm d}t' = 0 \,\, ,
\end{equation}
\noindent where $M_q (t)$ is a correlation function of
fluctuating forces \cite{Hansen86}. The kernel $M_q (t)$ can be split into a
regular part $M_q^{\rm reg} (t)$, dealing with conventional
liquid--state dynamics, and a contribution $\Omega_q^2 m_q (t)$,
dealing with slowly fluctuating forces due to sluggishly moving
structure: $M_q (t) = M_q^{\rm reg} (t) + \Omega_q^2 m_q (t)$.
Normal--state dynamics would then be obtained from the equation
${\cal D} \Phi_q (t) = 0$, where the abbreviation is used:
\begin{equation}
\label{A3}
{\cal D} \Phi_q (t) = \partial_t^2 \Phi_q (t) +
\Omega_q^2 \Phi_q (t) + \int_0^t M_q^{\rm reg} (t-t')
\partial_{t'} \Phi_q (t') {\rm d}t' \,\, .
\end{equation}
\noindent The subtleties of MCT are due to the approximation of
$m_q (t)$ as mode--coupling functional ${\cal F}_q$:
\begin{equation}
\label{A4}
m_q (t) = {\cal F}_q (\Phi_k (t)) = \Sigma_{kp} V_{q,kp}
\Phi_k (t) \Phi_p (t) \,\, .
\end{equation}
\noindent Here the positive coefficients $V_{q,kp}$ are given in
terms of $S_q$, i.e. by the equilibrium structure. The MCT
equations of motion are
\begin{equation}
\label{A5}
{\cal D} \Phi_q (t) + \Omega_q^2 \int_0^t m_q (t-t')
\partial_{t'} \Phi_q (t') {\rm d}t' = 0 \,\, .
\end{equation}
\noindent The theory is specified by $S_q$ and $M_q^{\rm reg}
(t)$ where neither quantity reflects glass--dynamics anomalies.
The Eqs. \gl{A3}--\gl{A5} are regular and $\Omega_q^2, V_{q,kp}, M_q^{\rm
reg} (t)$ depend smoothly on control parameters. For mathematical convenience 
the wave-vector
moduli are discretisized to a set of $M$ values up to some cutoff
$q_{\max}$. Thereby MCT deals with $M$ nonlinear integro--differential 
equations of the Volterra type, which are coupled
via the functional ${\cal F}_q$. A review of the derivation of the MCT
equations, 
in particular of the explicit form of $V_{q,kp}$, and citations
of the original papers can be found in Ref. \cite{Goetze95}.

Let us consider the hard-sphere system (HSS) as main example for the
following demonstration of our results. Its equilibrium
state is controlled by the packing fraction $\varphi = \pi
nd^3/6$. As unit of length the particle diameter $d$ is
chosen. Wave vectors are considered up to the cutoff $q_{\max} =
40$ and we use $M=100$ equally spaced $q$ values. The structure factor is evaluated within the
Percus--Yevick theory \cite{Hansen86}. Two models for the regular dynamics
shall be studied. In the first one the regular kernel is dropped,
so that
\begin{mathletters}
\begin{equation}
\label{A6a}
{\cal D}^{(1)} \Phi_q (t) = \partial_t^2 \Phi_q (t) + 
\Omega_q^2 \Phi_q (t) \,\, .
\end{equation}
\noindent This is a model with Newtonian dynamics where the transient deals
with oscillations. The second
model is obtained by dropping inertia effects and replacing the
regular kernel by a $q$--independent friction term $\nu$. This is a model for
a colloid where the transient deals with relaxators
\begin{equation}
\label{A6b}
{\cal D}^{(2)} \Phi_q (t) = \nu \partial_t \Phi_q (t) 
+ \Omega_q^2 \Phi_q (t) \,\, .
\end{equation}
\end{mathletters}

Another set of examples shall be formulated for a so--called schematic model
(SM), dealing with a single correlator $\Phi (t)$ only \cite{Goetze84}. 
The mode--coupling 
functional is specified as ${\cal F} (\Phi (t)) = v_1
\Phi (t) + v_2 \Phi (t)^2$, where $v_1 \geq 0 \,, \, v_2 \geq 0$
denote coupling constants. Despite its apparent simplicity, this $M=1$
model reproduces some generic features of the MCT so faithfully, that it
has been used as basis for a quantitative description of the
evolution of anomalous dynamics  of glycerol within the full
gigahertz band \cite{Franosch97a}. 
For the regular motion we again consider  the
oscillator model, Eq. \gl{A6a}, and the relaxator model, Eq. \gl{A6b}.
In addition we also study  a soft-mode contribution to the regular
dynamics. Here the transient, described by  ${\cal D}= {\cal D}^{(3)}$, is given by a
regular kernel in Eq. \gl{A3}, which obeys the oscillator equation
\begin{equation}
\label{A7}
(\partial_t^2 + \nu_0 \partial_t + \Omega_0^2)
M^{\rm reg} (t) = 0 \, ,
\end{equation}
\noindent with initial conditions $M^{\rm reg} (0) = c_0 \, ,
\partial_t M^{\rm reg} (0) = 0$. Choosing a low soft-mode
frequency $\Omega_0$ we shall
investigate the interference of slow regular dynamics with
anomalous dynamics.

\section{Anomalous MCT dynamics}\label{sectiondrei}

The solutions $\Phi_q(t)$ of the MCT equations of motion decay to
zero for large times as expected for a liquid, provided control
parameters like $n$ or $1/T$ are smaller than some critical value
$n_c$ or $1/T_c$. However, for $n \geq n_c$ or $1/T \geq 1/T_c$
the solutions describe ideal glass states characterized by a
positive Debye--Waller factor $f_q = \Phi_q (t \to \infty)$
\cite{Goetze92}. For the HSS one finds the bifurcation point for $\varphi_c
\approx 0.52$ \cite{Bengtzelius84}, which is not too far from the experimental
value 0.58 \cite{Megen94b}. The evolution of the HSS dynamics for wave vector
$q = 10.6$, calculated for the oscillation transient \gl{A6a}, is
shown in Fig. 1; and Figs. 2 and 3 exhibit the corresponding
fluctuation spectra $\Phi_q'' (\omega)$ and susceptibility spectra
$\chi_q'' (\omega) = \omega \Phi_q'' (\omega)$, respectively. Here
$\Phi_q''(\omega) = \int_0^\infty \cos (\omega t) \Phi_q(t) {\rm d}t$ is the
Fourier cosine transform of the correlator. The wave vector $q=10.6$ is located
close to the first minimum of the structure factor.
For the SM the weak--coupling 
liquid regime is separated from the strong--coupling
glass regime in the $v_1-v_2$--plane by the parabola $v_1^c = (2
\lambda - 1)/ \lambda^2$, $v_2^c = 1/ \lambda^2$, $1/2 \leq
\lambda < 1$ \cite{Goetze84}. Figures 4--6 exhibit the evolution of the
dynamics for the soft--mode model, Eq. \gl{A7}, upon crossing the
transition line at $\lambda = 0.7$. The Figs. 1--6 exhibit slow,
control--parameter--sensitive, stretched dynamics. Obviously, the results for
the HSS are similar to the corresponding ones for the SM and this exemplifies
the relevance of the latter for a discussion of MCT findings. The shown
anomalous dynamics can be understood from the asymptotic
solutions of the MCT equations for long times and low frequencies
near the transition, as explained in the preceding literature,
e.g. in Ref. \cite{Franosch97}.

There is a qualitative difference for the short--time
dynamics of the HSS between the results shown in Fig. 1 and the
ones shown in Ref. \cite{Franosch97} for the dynamics calculated for the
relaxation transient, Eq. \gl{A6b}. The correlators of the second
model decrease monotonously while the correlators in Fig. 1 exhibit oscillations. These get more and more
damped if $\varphi$ increases towards $\varphi_c$. Increasing $\varphi$ above
$\varphi_c$, the oscillations   
become again rather pronounced in the glass state. These yield the
oscillation bumps for the spectra $\Phi_q'' (\omega)$ for $\omega>10$ in Fig. 2,
while the spectra for the colloid model decrease monotonously
with increasing frequency \cite{Franosch97}. Similar differences occur for
the SM. The oscillations in Fig. 4 are different from the ones
calculated in Ref. \cite{Goetze96b} for the transient (6a); and the
correlators for the relaxation transient, Eq. \gl{A6b}, do not show
oscillations at all \cite{Goetze95}. The fluctuation spectra in Fig. 5
exhibit two oscillation bumps for $\omega > 0.1$, the spectra calculated with
Eq. \gl{A6a} have one bump for the glass states \cite{Goetze96b}, and the ones
for the 
relaxation transient \gl{A6b} have no bumps at all \cite{Goetze95}. However, 
the long--time 
parts of the decay curves referring to the same mode--coupling functional
${\cal F}_q$ coincide for the various models for the transient. This
holds provided an appropriate overall shift of the curves
parallel to the logarithmic abscissa is done in the figures. The
statement is demonstrated in Figs. 7, 8 for the two HSS models and in
Figs. 9, 10 for the three mentioned SMs.

The findings in Figs. 7--10 suggest the following precise formulation of
the
first proposition. The transient dynamics determines a time scale
$t_0$ so that for $t \gg t_0$ the correlators can be written as
\begin{equation}
\label{A8}
\Phi_q (t) = F_q (t/t_0) \,\, .
\end{equation}
\noindent Here $F_q$ is independent of  the
transient as quantified in Eq. \gl{A3} by $\Omega_q$ and $M^{\rm reg}_q(t)$. The  long--time decay,
its sensitive dependence on control parameters and also its
dependence on $q$, is given by the master functions $F_q$, which are determined
by the mode--coupling functional ${\cal F}_q$. Of
course, the transient may also depend on control parameters, and
this leads to a smooth variation of $t_0$. This effect results in
the  parallel shift of the $n = 2,~ \epsilon < 0$--curves in
Fig. 7 and similar
 offsets of the $n = 1$--curves in Figs. 9, 10. The proposition is meant as asymptotic
result for the approach towards the critical point so that $t_0$
is the limit result at the bifurcation singularity.

\section{The regular part \lowercase{ $t_0$} of the relaxation scales}

The first step of the proof of the first proposition is based on
the leading and next--to--leading long--time expansion of the
critical correlators, i.e., of the MCT solutions at the
bifurcation point:
\begin{equation}
\label{A9}
\Phi_q (t) = f_q^c + h_q (t_0/t)^a
\{ 1 + [K_q + \kappa(a)] (t_0/t)^a \} \,\, .
\end{equation}
\noindent Here terms of order $(t_0/t)^{3a}$ have been dropped.
There are straightforward formulas for the evaluation of the
critical form factor $f_q^c > 0$, the critical amplitude $h_q >
0$, the correction amplitude $[K_q + \kappa(a)]$, and the critical
exponent $a,~ 0<a<0.5$, from the mode-coupling functional \cite{Franosch97}.
These parameters are equilibrium quantities. For the HSS one gets
$a = 0.312,~ f_{10.6}^c = 0.417 ,~ h_{10.6} = 0.642 ,~
K_{10.6} + \kappa(a) = - 0.185$. For the SM one finds $a = 0.327 
,~ f^c = 0.3  ,~ h = 0.7  ,~ K =0,~  \kappa(a) = 0.0528$. The transient,
no matter how complicated $M_q^{\rm reg} (t)$ in Eq. \gl{A3} may be,
merely enters via the scale $t_0$. Specializing Eq. \gl{A8} to the
critical point, one identifies $t_0$ with the scale in the
proposition. There are two diverging
time scales hidden in $\Phi_q (t)$, which govern its sensitive
control--parameter dependence \cite{Goetze92}. Formula \gl{A8} implies, that
$t_0$ enters these scales as a prefactor. The singular
part of the scales is determined by the master functions $F_q(\tilde{t})$.

The found results are demonstrated in Figs. 11 and 12. For all
models of Sec. \ref{sectionzwei}, Eq. \gl{A9} was matched to the numerical solution
at the critical point. This determines the values $t_0$, cited
in the figure captions. Notice in Fig. 12 that a larger percentage
of the decay of $\Phi (t)$ is described by the law \gl{A9} for the
relaxator model than for the oscillator models. 
Slow transient
oscillations can destroy the short--time part of the critical fractal
decay. Given an upper time cutoff $t_{\max}$, the transient
oscillations can be modeled   so that they  destroy the critical decay 
for $t \leq t_{\max}$
completely. However, the decay for $t > t_{\max}$ is robust.
One can construct models so that there is a fully 
developed $\alpha$ process, i.e., a
stretched decay of $\Phi_q (t)$ from $f_q^c$ to zero, without a
critical decay precursor.

\section{A discrete--dynamics model}\label{sectionfuenf}

The second step of the proof of Eq. \gl{A8} is based on the Fourier--Laplace 
transform of the equations of motion \gl{A3}, \gl{A5}: $\Phi_q
(\omega) = - 1/ [\omega - \Omega_q^2/[\omega + M_q^{\rm reg}
(\omega) + \Omega_q^2 m_q (\omega) ]]$. Here we use the convention for the
transform of some function $G(t)$ to $G(\omega)$: 
$G(\omega) = {\rm i} \int_0^\infty \exp({\rm i} z t) G(t) {\rm d} t,~ z=
\omega +{\rm i} 0$. At the transition point
one derives from Eq. \gl{A9} a divergent small--frequency mode--coupling 
kernel $m_q (\omega) =[- {\cal F}_q (f_k^c)/ \omega] +{\cal O}(1/\omega^{1-a})$. Because of continuity one concludes
that in the asymptotic limit of small frequencies
and small separations from the critical point, the regular
function $\omega + M_q^{\rm reg} (\omega)$ can be neglected in
comparison to $\Omega_q^2 m_q (\omega)$. Hence, in this limit the
correlators obey $m_q (\omega) - \Phi_q (\omega) = \omega m_q
(\omega) \Phi_q (\omega)$ \cite{Goetze92}. Let us assume that $\Phi_q (t)$
can be continued as a function $F_q (t)$ to small times so, that
the formulated equation holds for all times and frequencies:
$[N_q (\omega) - F_q (\omega)] / \omega = N_q (\omega) F_q
(\omega) \, , N_q (t) = {\cal F}_q (F_k (t))$. The details of the
continuation are of no concern, since we are not interested in
the short--time transient. Backtransformation yields the set of $M$
implicit functional equations for the $M$ functions $F_q (t)$:
\begin{mathletters}
\label{A10}
\begin{equation}
\label{A10a}
\int_0^t [N_q (t') - F_q (t')] {\rm d}t' =
\int_0^t N_q (t-t')  F_q (t') {\rm d}t' \,\, ,
\end{equation}
\begin{equation}
\label{A10b}
N_q (t) = {\cal F}_q (F_k (t)) \, , 
\qquad q = 1, \ldots , M \,\, .
\end{equation}
\end{mathletters}
\noindent Equations of a similar form have been studied
before in some different context \cite{Goetze90,Goetze96}, and we shall adopt
some of the tricks of the preceding work to deal with the present problem.
Notice that Eqs. \gl{A10} cannot define a time scale, since they are
scale invariant. With $F_q (t)$
also $F_q^x
(t) = F_q (x \cdot t)$ is a solution for all $x > 0,~ q = 1, \ldots , M$ .

The integrals in Eq. \gl{A10a} shall be written as Riemann sums
formed on a time grid of equal step size $\delta$. The sums shall be 
determined by the values of the functions in the middle of the
intervals
\begin{equation}
\label{A11}
g_q^{(i)} = F_q ((i + 1/2) \delta ) \, ,
\qquad i = 0,1, \ldots \, \, .
\end{equation}
\noindent The sums can be regrouped so that Eqs. \gl{A10} read
\begin{mathletters}
\label{A12}
\begin{equation}
\label{A12a}
g_q^{(m)} = {\cal I}_q (g_k^{(0)} , g_k^{(1)}, \ldots ,
g_k^{(m)}) \, ,
\end{equation}
where the functional ${\cal I}_q$ is given by
\begin{equation}
\label{A12b}
{\cal I}_q
= \left\{ (1-g_q^{(0)}) {\cal F}_q (g_k^{(m)}) +
\sum_{i=0}^{m-1} \left[{\cal F}_q (g_k^{(i)}) - g_k^{(i)}\right]
- \sum_{i=1}^{m-1}{\cal F}_q (g_k^{(m-i)}) g_q^{(i)} \right\} /
[1 + {\cal F}_q (g_k^{(0)})] \,\, .
\end{equation}
\end{mathletters}
\noindent Formula (12a) can be considered as an implicit
equation to determine $g_q^{(m)}$ in terms of the $g_k^{(i)}$ for
$i$ preceding $m$:
\begin{equation}
\label{A13}
g_q^{(m)} = {\cal T}_q (g_k^{(0)}, g_k^{(1)}, \ldots ,
g_k^{(m-1)}) \, \, .
\end{equation}
\noindent The explicit solution for $g_q^{(m)}$, i.e., the construction 
of the
functional ${\cal T}_q$, is done by the following procedure.
One defines a sequence of approximands $g_{q,\ell}^{(m)} \, ,
\ell = 0,1,\ldots$, so that $\lim_{\ell \to \infty}
g_{q,\ell}^{(m)} = g_q^{(m)}$. Here $g_{q,\ell + 1}^{(m)} = {\cal
I}_q (g_k^{(0)}, \ldots , g_k^{(m-1)}, g_{k,\ell}^{(m)})$, and
the
start is chosen as $g_{k,0}^{(m)} = g_{k}^{(m-1)}$. Notice that
the scale invariance is reflected by the fact, that the step size
$\delta$ does not occur in Eqs. \gl{A12}, \gl{A13}.

The result \gl{A13} can be interpreted as an iterated mapping with
memory or, because of Eq. \gl{A11}, as the definition of a discrete
dynamics with retardation. The  $M \cdot m$ numbers $g_k^{(i)} , k
= 1, \ldots, M,~i=0,\ldots,m-1$ determine the $M$ numbers $g_q^{(m)}$. 
A sequence of values
$g_q^{(i)}$ is created, once that the $M$ initial values
$g_q^{(0)}$
are specified. The arbitrariness of the scale is hidden in the
one of the choice of the initial condition.

By construction one expects that the sequence defines a solution
of Eqs. \gl{A10}. To show this explicitly one can read all equations
backwards. As an approximand for the $F_q(t)$, step functions are 
 defined by $F_q^\delta (t) =
g_q^{(i)}$ for $i < t/ \delta \leq (i+1)$. Obviously,
$\lim_{\delta \to 0} F_q^\delta (t) = F_q (t)$ solves Eqs. \gl{A10},
and these functions agree with the master functions in Eq. \gl{A8} for
$t$ large compared  to the time scale $t_0$. The limit $\delta \to 0$
for $t \geq t_0$ is trivially taken by using Eq. \gl{A11} only in the
limit $i \to \infty$. In practice, a large $i_0$ is chosen and
the $g_q^{(i)}$ for $i < i_0$ are considered as transient of the
discrete mapping. For the prescribed small step size $\delta,~
\delta i_0$ has to be smaller than the lower bound of the time
window to be studied. The found results are then independent of
the indicated details, up to an overall time scale $t_0$. To cope with the
stretching of the dynamics over many decades, we applied a decimation
procedure \cite{Goetze96}. 

The discrete dynamics and thus the master functions $F_q$ have
been constructed in Eqs. \gl{A11}, \gl{A12}, \gl{A13} solely in terms of the
mode--coupling functional ${\cal F}_q$. Hence the function $F_q$
in Eq. \gl{A8} is a quantity defined by equilibrium distributions.
This finishes the proof of the first proposition. Figures 13 and
14 exhibit the construction of the master functions $F_q(t)$ for
the HSS and the SM. The corresponding results are included as
dotted lines in Figs. 7--12.

Some side remark shall be added. Equations (10), specialized to
the functional at the critical point and complemented by the
initial condition $F_q (t=0) = f_q^c$, are equivalent to the
equation for the MCT $\alpha$--relaxation master function \cite{Goetze92}.
The specified iterated mapping used with initial condition
$g_q^{(0)} = f_q^c - \delta^* h_q,~\delta^* > 0,~\delta^* \rightarrow 0$, is a very
simple and most efficient algorithm for the evaluation of $F_q(t)$.

\section{Complete monotonicity}\label{sectionsechs}

Let us combine Eqs. \gl{A5}, \gl{A6b} to the MCT equations of motion for
$M$ relaxators, specified by the decay times $\tau_q =
\nu / \Omega_q^2$
\begin{equation}
\label{A14}
\tau_q \partial_t \Phi_q (t) + \Phi_q(t) + 
\int_0^t m_q (t-t') \partial_{t'} \Phi_q (t') {\rm d} t' = 0 \,\, .
\end{equation}
\noindent Suppose that the kernels $m_q(t)$ are given as a superposition 
of $L$ Debye
processes: $m_q^{(*)} (t) = \sum_{j=1}^L \mu_q^j \exp (-
\Gamma_q^j t)$, where $\mu_q^j > 0$ and the rates are labeled so that $0 <
\Gamma_q^1 < \Gamma_q^2 < \cdots < \Gamma_q^L$. The 
Fourier--Laplace transform $m_q^{(*)} (\omega)$ is a meromorphic
function
with $L$ simple poles at $- {\rm i} \Gamma_q^j$. Substitution of this
result in the Fourier--Laplace transform of Eq. \gl{A14} yields the
solution
$\Phi_q^{(*)} (\omega)$ as meromorphic function. Elementary
discussion brings out that $\Phi_q^* (\omega)$ has exactly  $(L+1)$ poles
which are located on the imaginary axis at, say, $(- {\rm i}
\gamma_q^j)$; the poles of $m_q^*(\omega)$ separate those of
$\Phi_q^*(\omega)$, i.e.,  $\Gamma_q^{j+1} > \gamma_q^j > \Gamma_q^j$ for
$j = 1, \ldots, L - 1 ,~ \gamma_q^L > \Gamma_q^L  ,~
\Gamma_q^1 > \gamma_q^0 > 0$. The residues $\rho_q^j$ of the
poles are positive. Hence the solution is a superposition of
$(L+1)$ Debye processes $\Phi_q^{(*)} (t) = \sum_{j=0}^L \rho_q^j \exp (-
 \gamma_q^j t)$.

Neglecting the kernel $m_q(t)$, the solutions of Eq. \gl{A14} are Debye
processes $\Phi_q^{(0)} (t) = \exp (- t/\tau_q)$. The relaxators
are coupled by the mode--coupling functional (4), which is a
polynomial with positive coefficients $V_{q,kp}$. Hence ${\cal
F}_q (\Phi_k^{(0)} (t)) = m_q^{(1)} (t)$ is a sum of a finite number,
say $L$, Debye functions. Substitution of this kernel into Eq.
\gl{A14}, leads to  a solution $\Phi_q^{(1)}(t)$, which according to the preceding paragraph is a sum of $(L+1)$ Debye contributions. This can be
used to define a new sum of Debye processes $m_q^{(2)} (t) = {\cal
F}_q (\Phi^{(1)} (t))$. Continuing one constructs a sequence of
approximands $m_q^{(n)}(t),~\Phi_q^{(n)} (t) ,~ n = 0,1,2, \ldots$ of the
type formulated in Eq. \gl{A1}. In Ref. \cite{Goetze95b} it was shown: the
sequence $\Phi_q^{(n)} (t)$ converges uniformly towards the
unique solutions of Eq. \gl{A14} and this solution is completely
monotone, i.e., it obeys $(- \partial / \partial t)^\ell \Phi_q
(t) > 0 \, , \ell = 0,1, \ldots~$.

According to Bernstein's theorem \cite{Feller71}, every completely monotone
function can be represented as Stieltjes integral
\begin{equation}
\label{A15}
\Phi_q (t) = \int_0^\infty \exp (- \gamma t) 
{\rm d} \alpha_q (\gamma)  \,\, ,
\end{equation}
\noindent where the measure $\alpha_q (\gamma)$ is an increasing
function of the rate $\gamma$. The theory of these integrals
implies the following. For every finite  interval of positive times $t,~ 0<t_1
\leq t \leq t_2 < \infty $ and
every positive 
error margin $\eta$, one can find a set of numbers $\rho_q^j > 0
\, , \gamma_q^j \geq 0 \, , j = 1, \ldots , N$, so that
\begin{equation}
\label{A16}
| \Phi_q (t) - \sum\nolimits_{j=1}^N \rho_q^j
\exp (- \gamma_q^j t) | < \eta  \,\, .
\end{equation}
\noindent The solutions of Eq. \gl{A14} can be approximated
arbitrarily well by a finite sum of Debye processes.

Because of Eq. \gl{A8} the solutions of any MCT model are given for $t
\gg t_0$ by the transient independent function $F_q (t/t_0)$. This
holds in particular for the solution of Eq. \gl{A14}. Choosing $t_1$
larger than $t_0$ one can replace $\Phi_q (t)$ by $F_q (t/t_0)$ in
Eq. \gl{A16}. Thereby one obtains the precise formulation of the
second proposition and its proof.

\section{Conclusions}

The dynamics of a classical system deals with the orbits in phase
space. Collision events and vibrations are the elementary bits
building the motion on microscopic scales. In dense liquids there
is the cage effect \cite{Balucani94}: the system gets trapped in phase space
pockets for long times. This implies a separation of a low--frequency 
contribution to the spectra from the normal--motion
band. This contribution is the anomalous dynamics dealing with
the motion from one pocket to the other. The first proposition
leads to the following picture. The anomalous dynamics reflects
the statistics of orbits in configuration space. After coarse
graining of the time over intervals of microscopic size, the
normal condensed--matter dynamics does not play a role anymore;
it merely sets the scale $t_0$ for the exploration of the potential
landscape in the high--dimensional configuration space. It does
not matter, for example, whether this exploration is done as
prescribed by Newton's equations of motion or by Brownian dynamics.
Consequently, the correlation functions for the coarse-grained
dynamics are governed, up to an overall time scale $t_0$, only by Boltzmann factors, i.e. by
equilibrium distribution functions.

The statistics of orbits is studied in the  theory of generalized 
Brownian motion \cite{Ito65}. From this theory one expects as generic results 
 time fractals like the one formulated in Eq. \gl{A9}. These
fractals result from the mapping of the orbits on the 
time axis as  achieved by correlation functions.
They reflect nontrivial Hausdorff dimensionalities of  sets defined
for return--time and waiting--time  distributions. MCT should be viewed as a
mathematical model allowing an explicit evaluation of such orbit
statistics via the master functions $F_q$ in Eq. \gl{A8} or via the
discrete--dynamics procedure of Sec. \ref{sectionfuenf}.

Debye's relaxation law is the paradigm for a dynamics without
memory. Conventionally, it is derived by assuming random forces
with a white--noise spectrum \cite{Hansen86,Balucani94}. 
The derived Eqs. \gl{A1}, \gl{A16}
formulate a similar picture for the anomalous dynamics in glass--forming 
systems. Memory effects are irrelevant for the coarse--grained 
orbits through the potential landscape. At the first
glance this finding appears as a contradiction to  MCT, which deals with  a retarded fluctuating force
via the integral term in the equation of motion \gl{A5}. Indeed, 
it is this memory
term, which renders the MCT bifurcation scenario so different
from what one knows for bifurcations of conventional dynamical systems. The
solution of the paradox was explained in the first paragraph of
Sec. \ref{sectionsechs}. A  white--noise spectrum is sufficient to produce the
Debye law, but it is not necessary. If the forces are
superpositions of $L$ Debye laws, their retarded influence leads to a
response, which is also a superposition of Debye laws albeit of $(L+1)$ terms. The MCT 
memory effects do not destroy the complete monotonicity, i.e., the possibility
for a representation as superposition of Debye processes. But the memory
effects 
lead necessarily to a distribution of the relaxation rates, i.e.
to relaxation stretching.

Our result \gl{A8} implies that the susceptibility spectra can be written as
$\chi_q''(\omega) = \hat{\chi}_q(\omega t_0)$. Here the master spectrum
$\hat{\chi}_q(\omega)$ is
given by the equilibrium structure and it can be evaluated from the Fourier
cosine transform $F_q''(\omega)$ of the master functions $F_q(t)$, discussed in
Sec. \ref{sectionfuenf}: $\hat{\chi}_q(\omega) = \omega F_q''(\omega)$. This
holds, provided the frequencies are sufficiently small compared to the
characteristic scale $1/t_0$, which is determined by the transient. This
conclusion is demonstrated in Fig. \ref{fig15} for the two HSS models. For
$n\geq 6$, i.e. for $|\epsilon| = |(\varphi-\varphi_c)/\varphi_c| \leq
10^{-2}$, and $\omega t_0 \leq 10^{-2.5}$ the spectra for the relaxator model
(full lines) agree with those for the oscillator model (dashed lines), and both
agree with the results obtained for the discrete--dynamics model (dotted
lines). Two features of our findings should be emphasized. Firstly, the full
lines represent  spectra for the relaxator model, but only the
part for $\log_{10} \omega t_0 \leq -0.5$ is structural relaxation as 
given by $\hat{\chi}_q(\omega
t_0)$. Due to mode--coupling effects there are non--Debye relaxation spectra,
which deal with crossover phenomena from transient  to structural--relaxation
dynamics. The corresponding crossover window is much larger for the oscillator
model; it extends over the window $-2.5 < \log_{10} \omega t_0 < -0.5$. Here
the dashed curves in Fig. \ref{fig15} can be described reasonably by an
effective power law $\chi_q''(\omega) = h_q^{\mbox{\small eff}} (\omega
t_0)^{a_{\mbox{\tiny eff}}}$, where the effective amplitude $h_q^{\mbox{\small eff}}$ and
the effective exponent $a_{\mbox{\small eff}}$ depend on the control parameter
$\varphi$. Secondly, the range of the applicability of Eq. \gl{A8} can be
extended by incorporating the smooth control--parameter dependence of $t_0$,
mentioned at the end of Sec. \ref{sectiondrei}. Thereby the $\Phi(t)$ versus
$\log_{10}(t/t_0)$ diagrams, shown e.g. for $\epsilon < 0,~ n=1$ in
Fig. \ref{fig9}, get a shift parallel to the abscissa so that they coincide
with the dotted line there. Similarly, negative shifts bring the low--frequency
spectra in Fig. \ref{fig15} for $\epsilon < 0,~n=4$ for the two HSS models on
the master spectra, and positive shifts do the same for the $\epsilon>0,~n=4$
results. The important point is, that the same shifts do the rescaling for all
wave vectors $q$ for the HSS.

The critical decay law \gl{A9} is the germ of all analytical
results derived within MCT. It was first measured for CKN and that by
inelastic neutron scattering \cite{Knaak88}, and by polarized as well as
depolarized--light--scattering spectroscopy \cite{Tao91}. Later this part
of the anomalous spectrum was also explored for   a
number of other systems like, for example, orthoterphenyl \cite{Petry91,Steffen92,Cummins97}. The function $f + h/t^a$ with $f > 0\, , h > 0$ and $0 < a <
1$ is completely monotone and thus it can be written as Eq. \gl{A1}.
The $1/t^a$--decay leads to a self--similar spectrum $\Phi''(\omega) \propto
\omega^{1-a}$, which does
not define a time scale. The time scales $1/ \gamma_q^j$ in Eq.
\gl{A1} merely reflect the dynamical window, within which the
critical law was studied. The representation of the critical
spectrum by Debye peaks \cite{Steffen92} is not an alternative, let alone  a
phenomenological theory, for the original formulation of the
discovery \cite{Knaak88,Tao91,Petry91}, rather it is a reformulation in 
a manner
suggested by Bernstein's theorem. We have shown in Sec. \ref{sectionsechs} 
that
not only the critical spectrum, but the entire low--frequency 
spectrum can be represented as sum of Debye
contributions. 

\bigskip

\acknowledgments{We thank H.Z. Cummins, M. Fuchs, and W. Kob for discussions
and constructive criticism on our manuscript. 
Our work was supported by Verbundprojekt BMBF 03GO4TUM.}

\bibliographystyle{prsty}

\begin{thebibliography}{10}

\bibitem{Wong76}
J. Wong and C.~A. Angell, {\em Glass: Structure by Spectroscopy} (Marcel
  Dekker, Inc., New York, 1976).

\bibitem{Li92}
G. Li {\it et~al.}, Phys. Rev. A {\bf 45},  3867  (1992).

\bibitem{Cummins93}
H.~Z. Cummins {\it et~al.}, Phys. Rev. E {\bf 47},  4223  (1993).

\bibitem{Megen94b}
W. van Megen and S.~M. Underwood, Phys. Rev. E {\bf 49},  4206  (1994).

\bibitem{Goetze92}
W. G{\"o}tze and L. Sj{\"o}gren, Rep. Prog. Phys. {\bf 55},  241  (1992).

\bibitem{Hansen86}
J.-P. Hansen and I.~R. McDonald, {\em Theory of Simple Liquids}, 2nd ed.
  (Academic Press, London, 1986).

\bibitem{Goetze95}
W. G{\"o}tze and L. Sj{\"o}gren, Transp. Theory Stat. Phys. {\bf 24},  801
  (1995).

\bibitem{Goetze84}
W. G{\"o}tze, Z. Phys. B {\bf 56},  139  (1984).

\bibitem{Franosch97a}
T. Franosch, W. G{\"o}tze, M.~R. Mayr, and A.~P. Singh, Phys. Rev. E {\bf 55},
  3183  (1997).

\bibitem{Bengtzelius84}
U. Bengtzelius, W. G{\"o}tze, and A. Sj{\"o}lander, J. Phys. C {\bf 17},  5915
  (1984).

\bibitem{Franosch97}
T. Franosch {\it et~al.}, Phys. Rev. E {\bf 55},  7153  (1997).

\bibitem{Goetze96b}
W. G{\"o}tze and L. Sj{\"o}gren, Chem. Phys. {\bf 212},  47  (1996).

\bibitem{Goetze90}
W. G{\"o}tze, J. Phys.: Condens. Matter {\bf 2},  8485  (1990).

\bibitem{Goetze96}
W. G{\"o}tze, J. Stat. Phys. {\bf 83},  1183  (1996).

\bibitem{Goetze95b}
W. G{\"o}tze and L. Sj{\"o}gren, J. Math. Analysis and Appl. {\bf 195},  230
  (1995).

\bibitem{Feller71}
W. Feller, {\em Introduction to Probability Theory}, 2nd. ed. (Wiley, New York,
  1971), Vol.~II.

\bibitem{Balucani94}
U. Balucani and M. Zoppi, {\em Dynamics of the Liquid State} (Clarendon Press,
  Oxford, 1994).

\bibitem{Ito65}
K. Ito and H.~P. McKean, {\em Diffusion processes and their sample paths}
  (Springer, Berlin, 1965).

\bibitem{Knaak88}
W. Knaak, F. Mezei, and B. Farago, Europhys. Lett. {\bf 7},  529  (1988).

\bibitem{Tao91}
N.~J. Tao, G. Li, and H.~Z. Cummins, Phys. Rev. Lett. {\bf 66},  1334  (1991).

\bibitem{Petry91}
W. Petry {\it et~al.}, Z. Phys. B {\bf 83},  175  (1991).

\bibitem{Steffen92}
W. Steffen, A. Patkowski, G. Meier, and E.~W. Fischer, J. Chem. Phys. {\bf 96},
   4171  (1992).

\bibitem{Cummins97}
H.~Z. Cummins {\it et~al.}, Prog. Theor. Phys. {\bf 126},  21  (1997).

\end{thebibliography}

%\end{multicols}

\clearpage
\begin{figure}[ht]
\caption[Figur eins]{\label{fig1}}
\noindent Density correlators for the HSS for wave vector
$q = 10.6$ calculated for the oscillator transient, Eq. \gl{A6a}. The
time unit is chosen so that $v/d=2.5$. Curve $c$ refers to the critical
packing fraction $\varphi = \varphi_c$. The uppermost curve is
calculated for $\varphi = 0.60$, and the other curves refer to
$(\varphi - \varphi_c) / \varphi_c = \epsilon = \pm 1/10^{n/3} 
,~ n = 0,1, \ldots$. The free--oscillator curve $\Phi_q (t) = \cos
\Omega_q t$, which refers to $\epsilon = - 1$, is drawn only for
$\Omega_q t \leq 1.98$. The arrow marks the time $t_0= 0.00944$.
\end{figure}

\bigskip
\begin{figure}[ht]
\caption[Figur zwei]{\label{fig2}}
\noindent Density fluctuation spectra for the results of
Fig. 1.
\end{figure}

\bigskip

\begin{figure}[ht]
\caption[Figur drei]{\label{fig3}}
\noindent Susceptibility spectra for the results of Fig.
1.
\end{figure}

\bigskip

\begin{figure}[ht]
\caption[Figur vier]{\label{fig4}}
\noindent Correlators for the SM with soft--mode transient
according to Eq. \gl{A7} with parameters $\nu_0 = 0.2 \Omega,~\Omega_0=0.5
\Omega,~c_0=0.2$. The time unit is 
chosen so, that $\Omega = 1$. Curve $c$ refers to the critical
point $v_1^c = (2 \lambda - 1) / \lambda^2  ;~ v_2^c =
1/\lambda^2  ,~ \lambda = 0.7$. The others are calculated for
$v_{1,2} = v_{1,2}^c (1 + \epsilon)  ,~ \epsilon = \pm 1/4^n  ,~ n =
0,1,\ldots$. The arrow marks the time $t_0=0.0649$. 
\end{figure}

\bigskip

\begin{figure}[ht]
\caption[Figur fuenf]{\label{fig5}}
\noindent Fluctuation spectra for the results of Fig. 4.
\end{figure}

\newpage
\bigskip

\begin{figure}[ht]
\caption[Figur sechs]{\label{fig6}}
\noindent Susceptibility spectra for the results of Fig.
4.
\end{figure}

\bigskip

\begin{figure}[ht]
\caption[Figur sieben]{\label{fig7}}
\noindent Curves (1) are $\epsilon < 0$ results from Fig. 1 replotted
with scale $t_0 = 0.00944$. Curves (2) are the corresponding results
for the relaxation transient, Eq. \gl{A6b}, replotted from
Ref. \cite{Franosch97} 
with scale $t_0 = 0.425$. The dotted lines (3) are the 
discrete--dynamics--model results from Fig. 13, replotted with $t_0=267 \delta$. The
labels $n$  
are set as in Fig. 1. Curves for successive values of $n$ are shifted
horizontally by two decades in order to avoid overcrowding. 
\end{figure}

\bigskip

\begin{figure}[ht]
\caption[Figur acht]{\label{fig8}}
\noindent The analogous results as shown in Fig. 7 but for $\epsilon > 0$. 
\end{figure}

\bigskip

\begin{figure}[ht]
\caption[Figur neun]{\label{fig9}}
\noindent Curves (3) are $\epsilon < 0 $ results from Fig. 4 replotted
with scale $t_0 \Omega= 0.0649$. Curves (1) (and (2)) are  the
corresponding results calculated for the oscillation transient,
Eq. \gl{A6a} (relaxation transient, Eq. \gl{A6b}) replotted from Ref.
\cite{Goetze96b} (Ref. \cite{Goetze95}) with scales $t_0 \Omega = 0.0440$ $(t_0 \nu = 0.150)$.
The dotted lines (4) are discrete--dynamics--model results from Fig.
14, replotted with scale $t_0=26.2 \delta$. The labels $n$ are set as in
Fig. 4. Curves for successive values of $n$ are shifted horizontally by two
decades in order to avoid overcrowding.
\end{figure}

\bigskip

\begin{figure}[ht]
\caption[Figur zehn]{\label{fig10}}
\noindent The analogous results as shown in Fig. 9 but for $\epsilon > 0$.
\end{figure}

\clearpage

\begin{figure}[ht]
\caption[Figur elf]{\label{fig11}}
\noindent Density correlators of the HSS for $q = 10.6$
and the critical value $\varphi_c$ for the packing fraction.
Curve (1) is replotted from Fig. 1 with $t_0 = 0.00944$; curve (2) refers
to the relaxation transient, Eq. \gl{A6b}, and it is replotted from
Ref. \cite{Franosch97} with $t_0 = 0.425$. Curve (3) refers to the
discrete dynamics--model and is replotted from Fig. 13 with $t_0=267 \delta$. 
Curve (5) 
exhibits  the asymptotic expansion law, Eq. \gl{A9}, and the curve (4)
is the leading contribution to Eq. \gl{A9}: $\Phi_q(t) = f_q^c + h_q
(t_0/t)^a$. 
\end{figure}

\bigskip

\begin{figure}[ht]
\caption[Figur zwoelf]{\label{fig12}}
\noindent Decay curves for the SM for the critical point
$\lambda = 0.7$. Curve (1) refers to the oscillation model, Eq.
\gl{A6a}; it is taken from Ref. \cite{Goetze96b} and replotted with $t_0 \Omega =
0.0440$. Curve (2) refers to the relaxation model, Eq. \gl{A6b}; it is
taken from Ref. \cite{Goetze95} and replotted with $t_0 \nu = 0.1498$. Curve (3)
refers to the soft--mode model; it is replotted from Fig. 4 with
$ t_0 \Omega= 0.0649$. Curve (4) refers to the 
discrete--dynamics model and it is replotted from Fig. 14 with $t_0= 26.2\delta$. The
dashed line (5) exhibits  the asymptotic expansion law, Eq. \gl{A9}.
\end{figure}

\bigskip

\begin{figure}[ht]
\caption[Figur dreizehn]{\label{fig13}}
\noindent Density correlators of the HSS for wave vector
$q = 10.6$ calculated for the discrete--dynamics model with initial condition
$g_q^{(0)} = 10$. The time unit is chosen
so that $\delta=1$ in Eq. \gl{A11}.
The labeling of the curves is done as in Fig. 1. The arrow marks the time $t_0
=267$. 
\end{figure}

\bigskip

\begin{figure}[ht]
\caption[Figur vierzehn]{\label{fig14}}
\noindent Correlators of the SM for the 
discrete dynamics with $g^{(0)} = 10$. 
The time unit is chosen so that $\delta=1$ in Eq. \gl{A11}.
The labeling of the curves is
done as in Fig. 4. The arrow marks the time $t_0 =26.2$.
\end{figure}

\clearpage

\begin{figure}[ht]
\caption[Figur fuenfzehn]{\label{fig15}}
\noindent Susceptibility spectra for the HSS. The labeling of the curves is
done as in Fig. \ref{fig1}. The full lines are calculated for the relaxator
transient, Eq. \gl{A6b}, the dashed lines for the oscillator model
(Eq. \gl{A6a}), and the dotted results refer to the discrete dynamics model of
Sec. \ref{sectionfuenf}. The scales $t_0$ are listed in the caption of Fig. \ref{fig7}.
\end{figure}

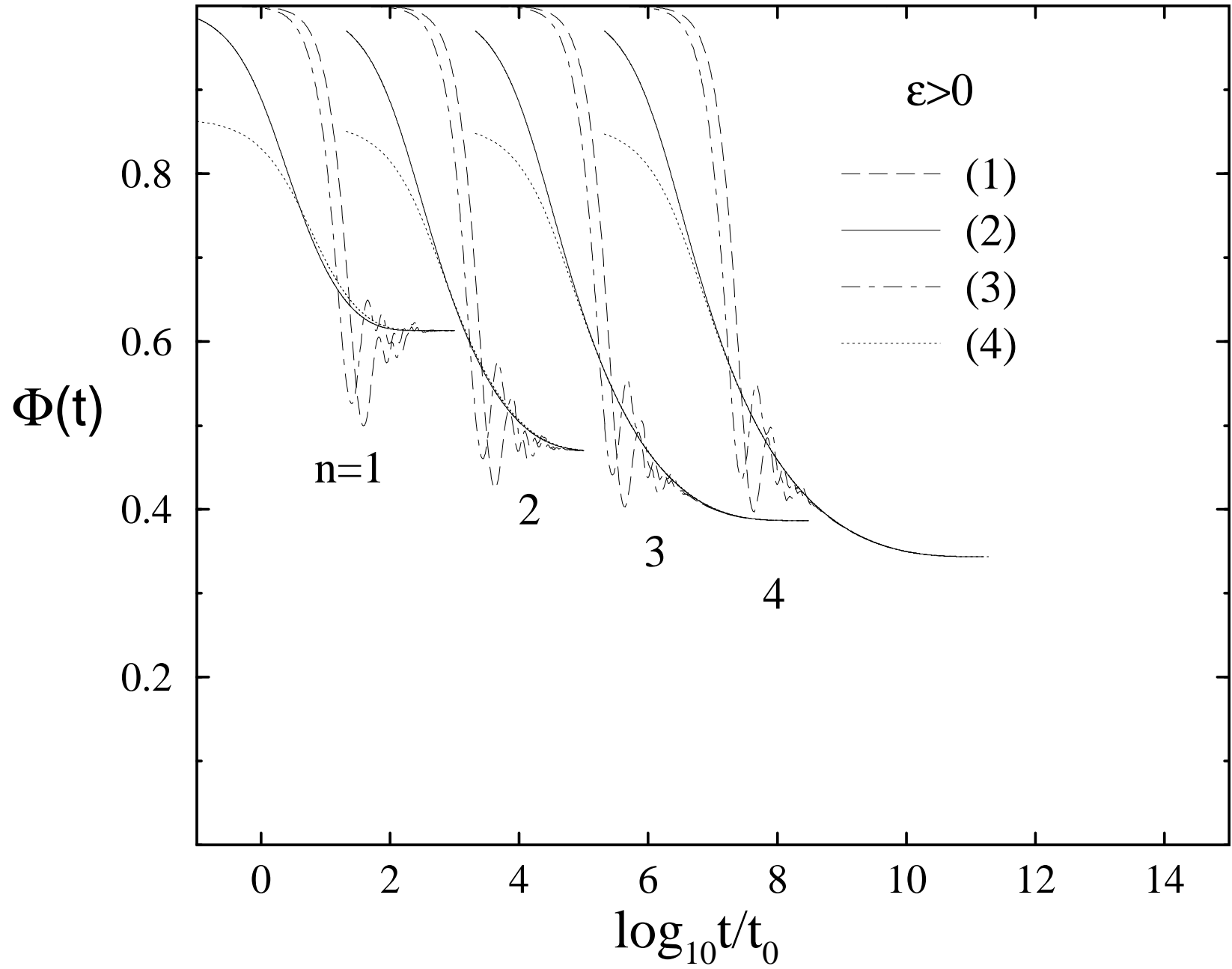
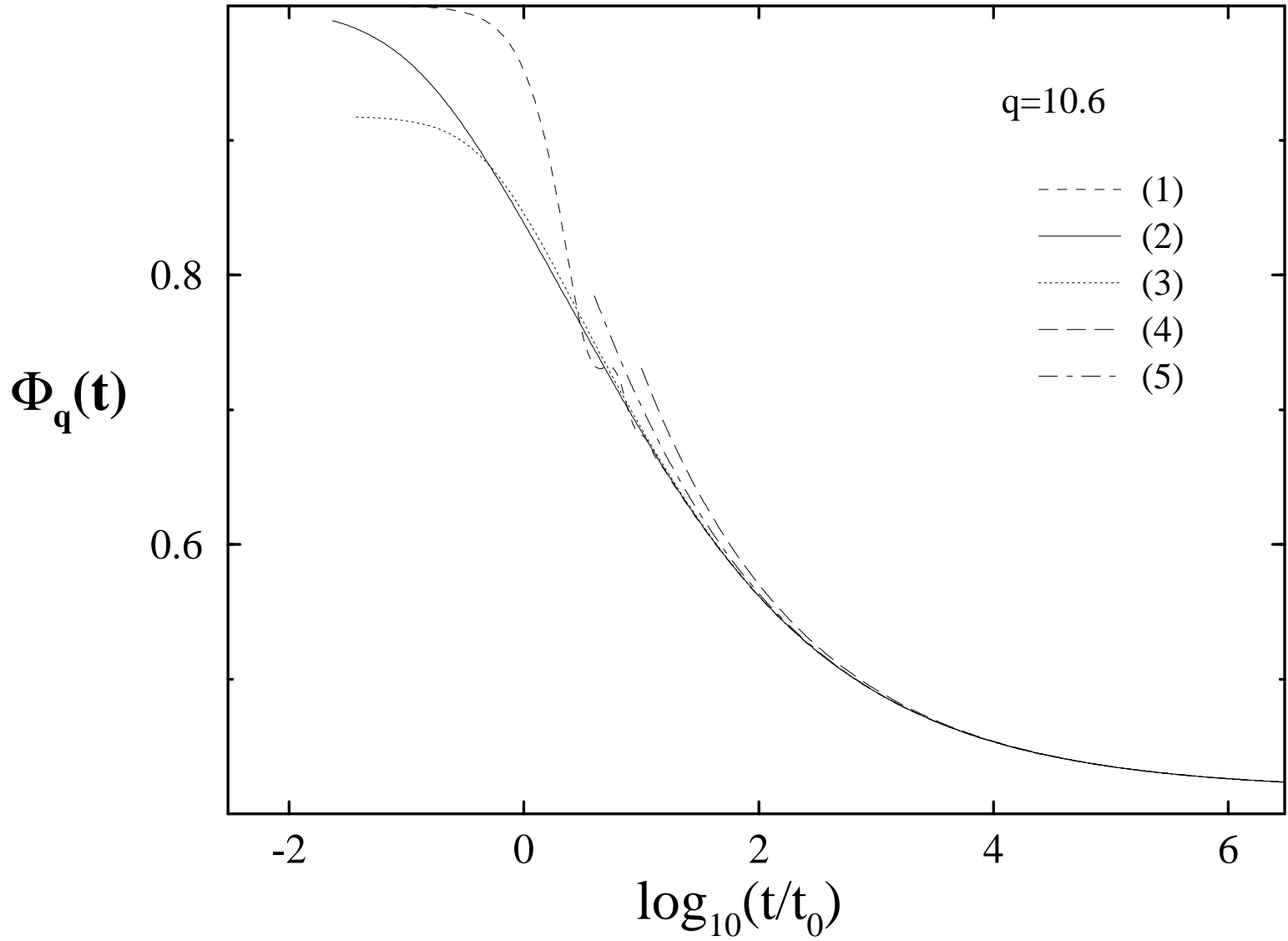
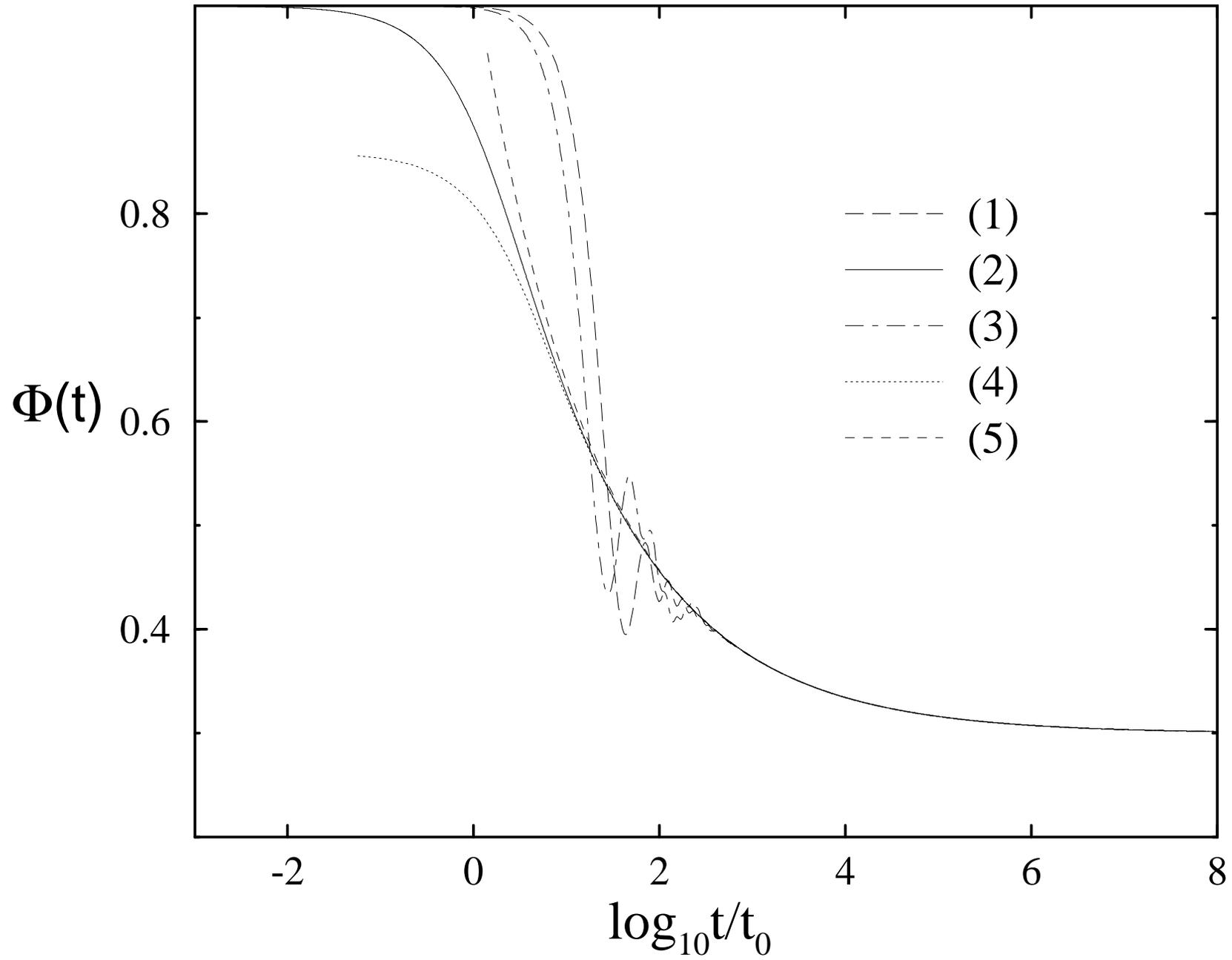
\end{document}